\documentstyle[12pt,aaspp4,psfig]{article}
 
\def\crad{c_{\rm r}}
\def\cgas{c_{\rm g}}
\def\vaz{v_{{\rm A}z}}
\def\vaphi{v_{{\rm A}\phi}}
\def\msun{{\rm M}_\odot}
\def\ledd{{\rm L}_{\rm Edd}}
\def\spose#1{\hbox to 0pt{#1\hss}}
\def\lta{\mathrel{\spose{\lower 3pt\hbox{$\mathchar"218$}}
     \raise 2.0pt\hbox{$\mathchar"13C$}}}
\def\gta{\mathrel{\spose{\lower 3pt\hbox{$\mathchar"218$}}
     \raise 2.0pt\hbox{$\mathchar"13E$}}}
\font\syvec=cmbsy10                        
\font\gkvec=cmmib10                         
\def\bnabla{\hbox{{\syvec\char114}}}       
\def\bphi{\hbox{{\gkvec\char30}}}          

\begin{document}
 
\title{Local Dynamical Instabilities in Magnetized,
Radiation Pressure Supported Accretion Disks}
 
\author{Omer Blaes and Aristotle Socrates}
\affil{Department of Physics, University of California, Santa Barbara,
CA 93106}

\begin{abstract}
We present a general linear dispersion relation which describes the
coupled behavior of magnetorotational, photon bubble, and convective
instabilities in weakly magnetized, differentially rotating accretion
disks.  We presume the accretion disks to be geometrically thin and
supported vertically by radiation pressure.  We fully incorporate the
effects of a nonzero radiative diffusion length on the linear modes.
In an equilibrium with purely vertical magnetic field, the vertical
magnetorotational modes are completely unaffected by compressibility,
stratification, and radiative diffusion.  However, in the presence of
azimuthal fields, which are expected in differentially rotating flows,
the growth rate of all magnetorotational modes can be reduced
substantially below the orbital frequency.  This occurs if diffusion
destroys radiation sound waves on the length scale of the instability,
and the magnetic energy density of the azimuthal component exceeds
the non-radiative thermal energy density.  While sluggish in this case,
the magnetorotational instability still persists and will still tap the
free energy of the differential rotation.  Photon bubble instabilities
are generically present in radiation pressure dominated flows where
diffusion is present.  We show that their growth rates are limited
to a maximum value which is reached at short wavelengths where the
modes may be viewed as unstable slow magnetosonic waves.  We also
find that vertical radiation pressure destabilizes upward propagating
fast waves, and that Alfv\'en waves can be unstable.  These
instabilities typically have smaller growth rates than the photon
bubble/slow modes.  We discuss how all these modes behave
in various regimes of interest, and speculate how they may affect
the dynamics of real accretion disk flows.
\end{abstract}

\keywords{accretion, accretion disks --- black hole physics ---
instabilities --- MHD}
 
\section{Introduction}

The physical state of the radiation pressure dominated, innermost regions
of accretion disks around black holes has been uncertain ever since the
early days of accretion disk theory.  Standard alpha disk models in which
the viscous stress is assumed to scale with the radiation pressure are
subject to thermal and viscous instabilities (Lightman \& Eardley 1974,
Shakura \& Sunyaev 1976).   These instabilities are sensitive to the
assumed prescription for the anomalous viscosity (e.g. Piran 1978),
so it remains unclear whether or not
they are actually present in real flows.  If they are, the accretion flow
may adopt a radically different state from that usually envisaged in thin
accretion disk theory, such as the multiphase equilibrium recently proposed
by Krolik (1998).

In addition to these secular instabilities, dynamical instabilities also
exist.  First and foremost, a differentially rotating flow with a negative
angular velocity gradient and an initially
weak magnetic field is unstable to the magnetorotational instability (MRI,
Balbus \& Hawley 1991).  The turbulence resulting from this instability
is currently the most plausible mechanism known for generating the
anomalous viscosity required in accretion disk models.  Dissipation of
magnetohydrodynamic waves excited by this turbulence by photon diffusion
and photon viscosity has recently been examined by Agol \& Krolik (1998).
Photon diffusion might also affect the linear development of the instability
itself in a laminar, radiation pressure dominated flow, but this issue
has not been examined previously.

Gammie (1998) has suggested that the overstable photon bubble modes discussed
by Arons (1992) in the context of X-ray pulsars also exist in radiation
dominated accretion flows in general.  However, his instability
analysis was limited to studying a static equilibrium where the effects
of differential rotation were entirely neglected.
Pietrini \& Krolik (2000) have recently investigated convective
instabilities in unmagnetized, radiation pressure dominated, differentially
rotating flows.
They assumed a constant vertical density profile, which necessarily leads
to an unstable entropy gradient.  Whether or not such unstable gradients
exist in reality depends on the vertical dissipation profile, which
in turn depends on the structure of the MHD turbulence.

These dynamical instabilities may all play a role
at some level in the dynamics and thermodynamics of the radiation pressure
dominated portion of accretion disks.  It is the purpose of this paper
to provide a unified description of all three by deriving a general
linear dispersion relation which incorporates them all.  Such an analysis
will hopefully prove to be a useful guide to numerical simulations which
explore the nonlinear development of these instabilities and their effects
on the resulting turbulent state of the inner accretion disk.

This paper is organized as follows.  In section 2 we discuss our basic
equations and assumptions.  Then in section 3 we focus exclusively on
the MRI by deriving its dispersion relation in the absence of
vertical stratification so that the other instabilities are suppressed.
We are therefore able to study just the effects of photon diffusion on
what is probably the most important of these instabilities.  In section
4 we then present the full dispersion relation for a vertically stratified,
differentially rotating medium, and discuss its solutions.
We discuss the relevance of these solutions to astrophysical
accretion disks in section 5 and then summarize our conclusions in section 6.

\section{Equations and Assumptions}

In Eulerian coordinates in an inertial frame, the equations of radiation MHD
which we use in this paper are
\begin{equation}
{\partial\rho\over\partial t} +{\bnabla\cdot}(\rho{\bf v})=0,
\end{equation}
\begin{equation}
\rho\left({\partial{\bf v}\over\partial t}+{\bf v\cdot\bnabla v}\right)
=-{\bnabla}p-\rho{\bnabla}\Phi+{1\over4\pi}({\bf\bnabla\times B}){\bf\times B}+
{\kappa_{\rm es}\rho\over c}{\bf F},
\label{gasmom}
\end{equation}
\begin{equation}
{\partial p\over\partial t}+{\bf v\cdot\bnabla}p={\gamma p\over\rho}
\left({\partial \rho\over\partial t}+{\bf v\cdot\bnabla}\rho\right),
\label{dsdt}
\end{equation}
\begin{equation}
{\partial E\over\partial t}+{\bf v\cdot\bnabla}E+{4\over3}E{\bf \bnabla\cdot v}
= -{\bf \bnabla\cdot F},
\label{dedt}
\end{equation}
\begin{equation}
{\bf F}=-{c\over3\kappa_{\rm es}\rho}{\bnabla}E,
\label{fdiff}
\end{equation}
and
\begin{equation}
{\partial{\bf B}\over\partial t}={\bnabla\times}({\bf v\times B}).
\label{dbdt}
\end{equation}
Here $\rho$ is the fluid mass density, ${\bf v}$ is the velocity, $p$ is the
gas pressure, ${\bf B}$ is the magnetic field, $\gamma$ is the gas adiabatic
index, $E$ is the radiation energy density, ${\bf F}$ is the
radiative flux, and $c$ is the speed of light.
We have followed the notation of Stone, Mihalas,
\& Norman (1992) with regard to variables associated with the radiation field.
In order to facilitate comparison with the work of Gammie (1998), we note that
his $p_g$ is our $p$, his $J$ is $cE/4\pi$, and his ${\bf H}$ is
${\bf F}/4\pi$.

We neglect the self-gravity of the fluid, and
assume that the gravitational potential $\Phi$ originates from the central mass.
We completely neglect relativistic effects.
We also assume a pure electron scattering opacity $\kappa_{\rm es}$, neglecting
absorption opacity entirely.  This should be a good first approximation to
the inner parts of accretion disks around black holes which are electron
scattering dominated.  Note, however, that this has the consequence that
the gas and radiation are coupled together purely by momentum exchange.
They do not exchange heat, which is why the gas energy equation (\ref{dsdt})
simply reduces to the condition for adiabatic flow.  This approximation still
recovers the basic MRI and photon bubble instabilities, and so we believe
captures the physics that we wish to explore here.

We have closed the radiation moment equations by assuming that the radiation
field is close to isotropic, so that the stress tensor is diagonal with
elements given by one third the radiation energy density.  We are therefore
neglecting the effects of photon viscosity considered for example by
Agol \& Krolik (1998).
The radiation momentum equation (\ref{fdiff}) deserves special comment.  We
have assumed a simple diffusion form, and have in particular neglected
time derivative and velocity-dependent terms that, e.g. Stone et al. (1992)
and Gammie (1998) have retained.  The full equation is
\begin{equation}
{1\over c^2}\left({\partial{\bf F}\over\partial t}+{\bf v\cdot\bnabla F}
+{\bf F\bnabla\cdot v}\right)=-{1\over3}{\bnabla}E-
{\kappa_{\rm es}\rho\over c}{\bf F}.
\label{dfdt}
\end{equation}
We are interested in perturbations with wavelengths $\lambda$ smaller than
the vertical disk scale height $H$, and angular frequencies $\omega$ or growth
rates which are of order the local angular velocity of the disk $\Omega$
(MRI, Balbus \& Hawley 1991) or possibly a factor $(2\pi H/\lambda)^{1/2}$
larger (photon bubbles, Gammie 1998).  All the terms on the left hand side of
equation (\ref{dfdt}) will be negligible provided
\begin{equation}
{\omega\over\Omega}\ll\tau_H\left({c\over c_{\rm s}}\right)
\,\,\,\,{\rm and}\,\,\,\,{v\over c}\ll{\tau_H\over2\pi}\left({\lambda\over H}
\right),
\label{dfdtcond}
\end{equation}
where $\tau_H\equiv\kappa_{\rm es}\rho H$ is the scattering optical depth
across a scale height, and $c_s$ is the total (gas plus radiation
pressure) sound speed.

Well outside the innermost stable circular orbit, the scattering depth in
the inner, radiation pressure dominated zone of a standard $\alpha$-viscosity
disk (Shakura \& Sunyaev 1973)\footnote{It is almost certainly the turbulence
generated by the MRI which is responsible for ``viscous'' stresses in accretion
disks, and therefore a value of $\alpha$.  We are nevertheless using $\alpha$
to give us some handle on the ambient conditions in a fictitious, laminar
accretion disk prior to the development of any instability.  The reader should
therefore view these scalings with appropriate caution.}~
is given by
\begin{equation}
\tau_H\sim1\times\alpha^{-1}\eta\left({L\over\ledd}\right)^{-1}
\left({R\over r_{\rm g}}\right)^{3/2}.
\end{equation}
Here $\eta\sim0.1$ is the radiative efficiency, $L/\ledd$ is the
luminosity scaled with Eddington, and $r_{\rm g}$ is the gravitational radius
of the central object.  The scattering depth is therefore
always significantly greater than unity except perhaps in the innermost
parts of disks accreting near the Eddington limit.  Radiative equilibrium,
hydrostatic equilibrium, and the assumption that the viscous stress is
$\alpha$ times the total pressure leads to a ratio of the speed
of light to the sound speed $c_{\rm s}\simeq(4E/9\rho)^{1/2}$ given by
\begin{equation}
{c\over c_{\rm s}}\sim\alpha\tau_H.
\end{equation}
Hence conditions (\ref{dfdtcond}) are almost always satisfied in the interior of
accretion disks, provided we do not consider wavelengths that are extremely
small.  The only exception is for near-Eddington disks in the innermost few
gravitational radii, where our Newtonian treatment of the magnetohydrodynamics
breaks down anyway.  We therefore choose to neglect the left hand side of
equation (\ref{dfdt}), which therefore reduces to the diffusion equation
(\ref{fdiff}).  The fact that Gammie (1998) found flux limited
electromagnetic waves as solutions of his dispersion relation is a reflection
of the fact that he retained these terms.  In most cases, these terms have
negligible effect on the photon bubble modes, which are at much lower
frequency.\footnote{Indeed, we have numerically recomputed all the
results in this paper with the terms on the left hand side of equation
(\ref{dfdt}) included and found no significant differences for realistic
physical parameters.}
Note that equation (\ref{fdiff}) allows us to write the radiation pressure
very simply in terms of a gradient in the radiation energy density $E$,
so the momentum equation (\ref{gasmom}) becomes
\begin{equation}
\rho\left({\partial{\bf v}\over\partial t}+{\bf v\cdot\bnabla v}\right)
=-{\bnabla}\left(p+{1\over3}E\right)-
\rho{\bnabla}\Phi+{1\over4\pi}({\bf\bnabla\times B}){\bf\times B}.
\end{equation}

Eliminating ${\bf F}$ using equation (\ref{fdiff}), we may write the radiation
energy equation (\ref{dedt}) as
\begin{equation}
{\partial E\over\partial t}+{\bf v\cdot\bnabla}E+{4\over3}E{\bf \bnabla\cdot v}
={\bf \bnabla\cdot}\left({c\over3\kappa_{\rm es}\rho}{\bnabla}E\right).
\end{equation}
In contrast to the radiation momentum equation (\ref{dfdt}), the left hand
side of this equation can be neglected only if
\begin{equation}
{\omega\over\Omega}\ll{4\pi^2\over3\tau_H}\left({c\over c_{\rm s}}\right)\sim
{4\pi^2\over3}\alpha
\left({H\over\lambda}\right)^2\,\,\,\,{\rm and}
\,\,\,\,{v\over c}\ll{2\pi\over3\tau_H}\left({H\over\lambda}\right).
\label{diffcond}
\end{equation}
Because $\alpha$ may be quite small and $\tau_H$ is not very small, these
conditions may not be satisfied, so we must retain the left hand side of the
radiation
energy equation (\ref{dedt}).  In fact, it is neglecting the {\it right} hand
side of this equation (photon diffusion) which returns us to the standard ideal
MHD equations with gas and radiation pressure.  When the conditions
(\ref{diffcond}) start to become satisfied, photon diffusion starts
to modify standard MHD.  It is this effect which gives rise to photon bubbles,
and which can modify the MRI.  We therefore choose to retain the left hand side
of equation (\ref{dedt}), although the right hand side can be
dominant.

\subsection{Equilibrium}

We consider a narrow annulus of the disk which, in equilibrium, is stationary
and axisymmetric with a purely azimuthal flow velocity
${\bf v}=R\Omega\hat\phi$.  (We assume cylindrical polar coordinates
$\{R,\phi,z\}$ throughout this paper.)
Following the usual analysis of the MRI as well
as previous analyses of photon bubbles, we assume the existence of a locally
uniform magnetic field.  Because we are considering a differentially rotating
system, we assume this field has no radial component, which would otherwise be
sheared into a time-dependent azimuthal field.  Hence ${\bf B}=B_\phi\hat{\bphi}+
B_z\hat{\bf z}$.  With the exception of the external gravitational potential
$\Phi$, we assume that all radial gradients are much smaller than vertical
gradients of the same quantities, i.e. that the disk is geometrically thin.
The radiative diffusion equation (\ref{fdiff}) then implies that the equilibrium
radiative flux is mainly in the vertical direction, and thus the radial component
may be neglected.
We will allow for the possibility of radial fluid gradients creating
small departures from Keplerian flow by retaining a general form for the
angular velocity profile $\Omega(R)$ in our dispersion relation.

With these
assumptions, the only nontrivial equilibrium equations are those expressing
vertical hydrostatic balance,
\begin{equation}
-{1\over\rho}{\partial p\over\partial z}+{\kappa_{\rm es}\over c}F_z-g=0,
\label{hydrostat}
\end{equation}
radiative equilibrium,
\begin{equation}
{\partial F_z\over\partial z}=0,
\end{equation}
and radiation diffusion,
\begin{equation}
F_{z}=-{c\over3\kappa_{\rm es}\rho}{\partial E\over{\partial z}}.
\end{equation}
Here $g\equiv\partial\Phi/\partial z$ is the magnitude of the local vertical
gravitational acceleration produced by the tidal field of the central mass.

This equilibrium is characterized by a number of parameters which are important
in describing wave modes.  First are the gas and radiation sound
speeds, defined as
\begin{equation}
\cgas=\left({\gamma p\over\rho}\right)^{1/2}\,\,\,\,{\rm and}
\,\,\,\,\crad=\left({4E\over9\rho}\right)^{1/2},
\end{equation}
respectively.
Then there are the Alfv\'en speeds associated with each component of the
magnetic field,
\begin{equation}
\vaphi={B_\phi\over(4\pi\rho)^{1/2}}\,\,\,\,{\rm and}\,\,\,\,
\vaz={B_z\over(4\pi\rho)^{1/2}}.
\end{equation}
Internal hydrodynamic modes are driven by equilibrium gradients with associated
oscillation frequencies.  The square of the epicyclic frequency is given by
\begin{equation}
\kappa^2={1\over R^3}{d\over dR}(R^4\Omega^2).
\end{equation}
It is convenient to define the square of the Brunt-V\"ais\"al\"a frequency
separately in the gas and radiation as
\begin{equation}
N_{\rm g}^2=g\left[{1\over\rho\cgas^2}\left({\partial p \over\partial z}\right)
-{\partial\ln\rho\over\partial z}\right]
\end{equation}
and
\begin{equation}
N_{\rm r}^2=g\left[{1\over3\rho\crad^2}\left({\partial E \over\partial z}\right)
-{\partial\ln\rho\over\partial z}\right],
\end{equation}
respectively.  Then the square of the total Brunt-V\"ais\"al\"a frequency is
\begin{equation}
N^2=g\left[{1\over\rho(\cgas^2+\crad^2)}{\partial
\over\partial z}\left(p+{1\over3}E\right)-{\partial\ln\rho\over\partial z}
\right]={\cgas^2N_{\rm g}^2+\crad^2N_{\rm r}^2\over\cgas^2+\crad^2}.
\label{brunt}
\end{equation}
Finally, there is a characteristic length scale associated with radiative
diffusion, given by how far photons diffuse on an orbital time.  This
can be used to define a diffusion wavenumber as
\begin{equation}
k_{\rm diff}\equiv\left({3\kappa\kappa_{\rm es}\rho\over c}\right)^{1/2}.
\end{equation}

\subsection{Perturbation Equations}

Linearizing our equations of motion about the equilibrium, and assuming
a WKB space-time dependence $\propto\exp[i(\int k_RdR+\int k_zdz -\omega t)]$,
we arrive at the following perturbation equations:
\begin{equation}
-i\omega\delta\rho + i\rho {\bf k}\cdot\delta {\bf v} +
\delta v_z{\partial\rho\over\partial z}=0,
\label{dcont}
\end{equation}
\begin{equation}
-i\omega\rho\delta v_{R} -2\Omega\rho\delta v_{\phi}=
-ik_{R}\left(\delta p+{1\over3}\delta E\right)
-i{k_{R}\over 4\pi}[B_{\phi}\delta B_{\phi} +B_{z}\delta B_{z}]
+i{k_{z}\over 4\pi}B_{z}\delta B_{R},
\end{equation}
\begin{equation}
-i\omega\rho\delta v_{\phi} + {{\kappa}^2\over 2\Omega}\rho\delta v_{R}=
i{k_{z}\over 4\pi}B_{z}\delta B_{\phi},
\label{dvphi}
\end{equation}
\begin{equation}
-i\omega\rho\delta v_{z} = -ik_{z}\left(\delta p+{1\over3}\delta E\right)
-g\delta\rho-i{k_{z}\over 4\pi}B_{\phi}\delta B_{\phi},
\label{dvz}
\end{equation} 
\begin{equation}
-i\omega\delta p +\delta v_{z}{\partial p\over\partial z}=
-i\gamma p {\bf k}\cdot\delta{\bf v},
\label{dengas}
\end{equation} 
\begin{equation}
-i\omega\delta E+\delta v_{z} {\partial E\over \partial z}
+i{4\over3}E{\bf k}\cdot\delta {\bf v}=
-i{\bf k}\cdot\delta {\bf F},
\end{equation}
\begin{equation}
\delta F_R=-{ic\over 3\kappa_{es}\rho}k_R \delta E,
\end{equation}
\begin{equation}
\delta F_\phi=0,
\label{dfphi}
\end{equation}
\begin{equation}
\delta F_z=-F_z{\delta\rho\over\rho}-{ic\over 3\kappa_{es}\rho}k_z \delta E,
\label{dfz}
\end{equation}
\begin{equation}
-i\omega\delta B_{R}=ik_{z}B_{z}\delta v_{R},
\label{dbr}
\end{equation}
\begin{equation}
-i\omega\delta B_{\phi} - R{d\Omega\over dR}\delta B_{R}=
-iB_{\phi}{\bf k}\cdot\delta {\bf v}+ik_{z}B_{z}\delta v_{\phi},
\label{dbphi}
\end{equation}
and
\begin{equation}
-i\omega\delta B_{z}=-ik_{R}B_{z}\delta v_{R}.
\end{equation}
The radial and vertical induction equations together
guarantee that the divergence of the magnetic field perturbation vanish.
Hence one or the other can be replaced with
\begin{equation}
ik_R\delta B_R + ik_z\delta B_z = 0.
\label{divdb}
\end{equation}

Equations (\ref{dcont}) and (\ref{dengas})-(\ref{dfz}) can be solved to yield
a simple expression for the perturbed total pressure, viz.
\begin{equation}
\delta p+{1\over3}\delta E=\left(\cgas^2+{\omega\over\omega_{\rm diff}}\crad^2
\right)\delta\rho-{\rho\over g}\left(\cgas^2N_{\rm g}^2+
{\omega\over\omega_{\rm diff}}\crad^2N_{\rm r}^2\right)
{i\delta v_z\over\omega}-
{k_zF_z\over3\omega_{\rm diff}}{\delta\rho\over\rho},
\label{deltaptot}
\end{equation}
where
\begin{equation}
\omega_{\rm diff}\equiv\omega+i{k^2\over k_{\rm diff}^2}\kappa.
\end{equation}
Each term on the right hand side of equation
(\ref{deltaptot}) has a simple interpretation.  The first is the acoustic
pressure perturbation in response to rarefaction or compression.  The
second is the buoyancy force (note that $i\delta v_z/\omega$ is simply
the vertical Lagrangian displacement of a fluid element).  The last term
is the radiation pressure perturbation arising from the enhanced (diminished)
optical depth in compression (rarefaction) regions.  The middle term can
give rise to convection, while the last term is responsible for photon
bubble modes.

The twelve equations (\ref{dcont})-(\ref{dbphi}) and (\ref{divdb}) relate
twelve perturbation variables.  Eight of these equations have an explicit
$\omega$ dependence, and so we therefore expect an eighth order dispersion
relation.  However, as we show in Appendix A, one of these modes is in fact
a zero frequency solution which merely represents a perturbation to the
equilibrium.  The interesting physics therefore lies in a seventh order
dispersion relation.
In section 4, we will present this full
dispersion relation, but we will first discuss in section 3 the simpler case
of the MRI in a non-stratified equilibrium.

\section{The Magnetorotational Instability in the Absence of Stratification}

In the fictitious laminar flow we are considering, it is presumably the
MRI which generates the turbulence whose dissipation, combined with
cooling, sets up the thermal
profile which then might drive photon bubble and convective
modes.  We therefore begin by considering the MRI in isolation, but
in a flow which is still hot enough that radiation pressure dominates
gas pressure.  In the absence of equilibrium stratification,
the perturbation equations can be solved
to yield the following dispersion relation:

\begin{equation}
D_{\rm ms}D_{\rm BH}+k_z^2\vaz^2\vaphi^2\left(k_z^2R
{d\Omega^2\over dR}-k^2\tilde\omega^2\right)-{k_R^2\over k_z^2}
\tilde\omega^2\omega^4=0.
\label{nonstratdisp}
\end{equation}
Here
\begin{equation}
\tilde\omega^2\equiv\omega^2-k_z^2\vaz^2,
\end{equation}
\begin{equation}
D_{\rm ms}\equiv\omega^2-k_z^2\left(\cgas^2+{\omega\over
\omega_{\rm diff}}\crad^2+\vaphi^2\right)
\end{equation}
are magnetosonic dispersion terms,
and
\begin{equation}
D_{\rm BH}\equiv{k^2\over k_z^2}\tilde\omega^4-\kappa^2\tilde\omega^2-4
\Omega^2k_z^2\vaz^2
\end{equation}
is the MRI dispersion relation (if set to zero) in a non-stratified,
incompressible medium (Balbus \& Hawley 1991).

The dispersion relation (\ref{nonstratdisp}) describes how the MRI is coupled
to magnetosonic modes when allowance is made for compressibility and radiation.
Without radiation, it reduces to equation (64) of Blaes \& Balbus (1994)
for $k_R=0$, which is all they considered.
The conclusion of those authors remains valid here: provided
the equilibrium azimuthal magnetic field is {\it sufficiently} subthermal, with
respect to either the gas or radiation energy densities, then the MRI remains
essentially unchanged.
Mathematically, this is because the Alfv\'en speed is then much
smaller than the relevant sound speed.  For the MRI,
$|\omega|\sim\Omega\sim k\vaz$, and the dispersion relation reduces to 
\begin{equation}
\left(\cgas^2+{\omega\over
\omega_{\rm diff}}\crad^2\right)D_{\rm BH}=0,
\end{equation}
so that the MRI mode frequencies remain unchanged from their incompressible
form.  Physically, in subthermal magnetic fields, the relevant wave speed
for the MRI (the Alfv\'en speed) is much slower than the sound speed, so
the associated fluid motions are very nearly incompressible.

The question of how subthermal the azimuthal field must be in order not
to affect the MRI is determined by radiative diffusion.  From the
form of the dispersion relation (\ref{nonstratdisp}), it is clear that the MRI
will be unaffected provided
\begin{equation}
\vaphi^2\ll\cgas^2+\crad^2\left(1+{\Omega^4\over\vaz^4k_{\rm diff}^4}
\right)^{-1/2}.
\label{vaphicond}
\end{equation}
A sufficient condition for this to be true is that $\vaphi\ll\cgas$.  However,
in a radiation pressure dominated region of an accretion disk, it is
reasonable to consider azimuthal fields for which $\cgas<\vaphi<\crad$.
In this case the MRI can be modified if the characteristic wavenumber for
instability $\Omega/\vaz$ exceeds the diffusion wavenumber $k_{\rm diff}$,
i.e. photons have time to diffuse across a wavelength in an orbital time.
This reduces the effective sound speed, making the fluid more compressible.
The criterion that the MRI be unaffected is
more stringent: $\vaphi\ll\crad(k_{\rm diff}\vaz/\Omega)$.

Compressibility acts to reduce the growth rate of the MRI in the presence
of azimuthal fields, but it does not remove the instability (Blaes \& Balbus
1994).\footnote{We do not have a rigorous proof that the actual stability
criterion remains unchanged in our case.  However, we suspect that this is
true based on our numerical calculations illustrated in Fig. 1.  In addition,
the dispersion relation (\ref{nonstratdisp}) admits a zero frequency root
for $\omega$  when $k^2\vaz^2=-R(d\Omega^2/dR)$ (see section 4), which is
the condition
for the marginally stable wavenumber of the MRI in incompressible MHD
in the absence of stratification.}~
This is illustrated in Fig. 1, which depicts the maximum growth rate
as a function of azimuthal field strength and diffusion wavenumber.  We have
restricted consideration to vertical ($k_R=0$) modes, as these are expected
to have the fastest growth rates, and we consider an equilibrium with
negligible gas pressure ($\cgas=0$).  At high values
of $k_{\rm diff}\vaz/\Omega$, photon diffusion is negligible, and we recover
the MRI behavior for an ideal compressible fluid.  Subthermal azimuthal
fields then produce no effect on the growth rate of the instability, but when
the field starts to become thermal in strength, there is a modest reduction
in the growth rate.  For
$k_{\rm diff}\vaz/\Omega\lta1$, photon diffusion is important.  In this
regime, azimuthal fields can reduce the growth rate quite substantially if
$\vaphi\gta\crad(k_{\rm diff}\vaz/\Omega)$.  For the parameters chosen in the
figure, this corresponds to $\vaphi/\vaz\gta1$ for
$k_{\rm diff}\vaz/\Omega=0.1$.
Higher values of $k_{\rm diff}\vaz/\Omega$ (less diffusion) require higher
azimuthal field strengths to affect the instability.

We can estimate how small the MRI growth rate becomes when diffusion is
important by
setting $k_R=\crad=0$ in equation (\ref{nonstratdisp}) and then solving
it in the limit $\cgas\rightarrow0$.  The MRI then satisfies
\begin{equation}
\omega^2\simeq{k^2\cgas^2\vaz^2\left(R{d\Omega^2\over dR}+k^2\vaz^2\right)
\over\vaz^2\left(R{d\Omega^2\over dR}+k^2\vaz^2\right)+\vaphi^2
(\kappa^2+k^2\vaz^2)}.
\end{equation}
For $\vaphi\gg\vaz$, the unstable growth rate is therefore reduced to
$\sim\Omega(\cgas/\vaphi)$.

Note that regardless of photon diffusion, the MRI is virtually unaffected by
compressibility provided the field is sufficiently vertical, i.e.
$\vaphi\ll\vaz$ (e.g. the $\vaphi/\vaz=0.1$ curve shown in Fig. 1).
This is because a flow with a purely vertical
equilibrium field will have vertical ($k_R=0$) MRI modes identical to
the incompressible case, as the dispersion relation (\ref{nonstratdisp})
permits $D_{\rm BH}=0$ as a solution.

To summarize, the MRI is essentially unaffected by photon diffusion, unless
the azimuthal component of the field is comparable to or larger than the
vertical component, and the vertical component of the field is weak enough
that photon diffusion is important on the scale of an unstable wavelength
$\sim\vaz/\Omega$.  While the instability still persists, even subthermal
azimuthal field components can then
reduce the growth rate to $\sim\Omega(\cgas/\vaphi)$
if condition (\ref{vaphicond}) is violated.

\subsection{The Relation Between the MRI, Alfv\'en and Slow Waves}

In the next section we will examine stratified equilibria, which turn
out to be subject to a number of unstable modes which can couple to
each other in complicated ways.  To help disentangle this complexity, it is
helpful to briefly consider the relation between the MRI and classical
MHD waves.

Normally, in the absence of radiative diffusion, the MRI can
be viewed as a differential rotation driven destabilization of the
slow magnetosonic mode in a weakly magnetized medium (Balbus \& Hawley
1998).  The distinction between slow and Alfv\'en modes is delicate, however.
In the limit where the magnetic field becomes very weak, the slow magnetosonic
mode becomes incompressible and has a dispersion relation which is the same
as that of Alfv\'en waves.  It is in fact identical in this limit to the
``pseudo-Alfv\'en'' mode which exists in incompressible fluids.  The
prefix ``pseudo'' is used because, although the mode is degenerate with
the Alfv\'en mode, the eigenfunction is still different.  Alfv\'en waves have
velocity perturbations which are perpendicular to the plane of the wave vector
and equilibrium magnetic field.  Slow, or pseudo-Alfv\'en, waves have
velocity perturbations in the plane of the wave vector and equilibrium
magnetic field.  For example, if $B_\phi=0$ in our geometry, Alfv\'en
waves would involve only $\delta v_\phi$, whereas slow waves would have
nonzero $\delta v_R$ and $\delta v_z$.

When the Coriolis forces associated with rotation are included, radial
and azimuthal motions are coupled together, so the distinction between
slow and Alfv\'en modes is blurred further.  It is therefore perhaps
accidental that the stable modes to which the MRI converts are
normally the slow modes at high wavenumber.  Figure 2(a) illustrates this
fact in the case of negligible radiative diffusion.  However, when radiative
diffusion is important, the MRI can convert to the Alfv\'en wave [Fig. 2(b)]
or the slow wave [Fig. 2(c)].

This would only be of academic interest were it not for the fact that it
is the slow mode which continuously transforms at low
wavenumbers into the unstable photon bubble mode in static, stratified
equilibria (Gammie 1998).  The relationship between this instability and
the MRI in linear theory therefore depends on whether the Alfv\'en or slow
mode is more closely tied to the MRI.  Also, as we will show in the next
section, the Alfv\'en mode is itself unstable in stratified equilibria.

\section{The Dispersion Relation in the Presence of Vertical Stratification}

With considerable algebra, the general perturbation equations from section
2 can be used to derive the following seventh order dispersion relation:
\begin{eqnarray}
& &D_{\rm ms}D_{\rm BH}+k_z^2\vaz^2\vaphi^2\left(k_z^2R
{d\Omega^2\over dR}-k^2\tilde\omega^2\right)-{k_R^2\over k_z^2}
\tilde\omega^2\omega^4\nonumber\\
& &+\left(D_{\rm BH}-{k_R^2\omega^2\tilde\omega^2\over k_z^2}\right)
\left[-{k_zk_R^2F_z\over3\omega_{\rm diff}\rho}+
\left(g-i{k_z^2F_z\over3\omega_{\rm diff}\rho}\right){1\over\rho}
{\partial\rho\over\partial z}\right]\nonumber\\
& &+i2\Omega\omega k_zk_R\vaz\vaphi\left[-{k_zk^2F_z\over3\omega_{\rm diff}
\rho}-i{k_z^2F_z\over3\omega_{\rm diff}\rho}{1\over\rho}
{\partial\rho\over\partial z}-2ik_zg\right]\nonumber\\
& &+k_R^2\tilde\omega^2\left(\cgas^2N_{\rm g}^2+{\omega\over\omega_{\rm diff}}
\crad^2N_{\rm r}^2\right)
+{k_R^2k_z\tilde\omega^2F_z\omega^2\over3\omega_{\rm diff}\rho}
-{g\over\rho}{\partial\rho\over\partial z}k_R^2\vaphi^2\omega^2=0.
\label{stratdisp}
\end{eqnarray}
The first three terms are identical to the non-stratified dispersion relation
(\ref{nonstratdisp}) discussed in the last section.  All other terms
arise from the stratification.  Note that in order for the WKB approximation
to be consistent, we must have $|k_z|\gg|\partial\ln\rho/\partial z|$, so
some of the terms with explicit dependence on the vertical density gradient
should probably be taken to be negligible.  In a radiation pressure dominated
accretion disk, the WKB approximation also demands that
\begin{equation}
k\gg{g\over\crad^2}\simeq{\kappa_{\rm es}F_z\over c\crad^2}.
\label{wkbcond}
\end{equation}

The dispersion relation (\ref{stratdisp}) reduces to those of previous authors
under various limits.  First, in the incompressible limit where
$(\cgas^2+\omega\crad^2/\omega_{\rm diff})\rightarrow\infty$, we recover the
MRI dispersion relation in the presence of vertical stratification (Balbus
\& Hawley 1991), albeit slightly modified by radiative diffusion:
\begin{equation}
{k^2\over k_z^2}\tilde\omega^4-\left[\kappa^2+{k_R^2\over k_z^2}\left(
{\cgas^2N_{\rm g}^2+\omega\crad^2N_{\rm r}^2/\omega_{\rm diff}\over
\cgas^2+\omega\crad^2/\omega_{\rm diff}}\right)\right]
\tilde\omega^2-4\Omega^2k_z^2\vaz^2=0.
\end{equation}
In addition, if we first assume that gas pressure is negligible
($\cgas^2\rightarrow0$), adopt the ansatz that $\omega\sim k^{1/2}$
and then take the short wavelength limit $k\rightarrow\infty$, we obtain
two roots given by
\begin{equation}
\omega^{2}=-ig\left({k_zk_R^2\over k^2}\right)\left({B_z\over B}\right)^2.
\label{pbdisp}
\end{equation}
This is identical to the asymptotic photon bubble instability dispersion
relation obtained by Gammie (1998, eq. 41), adapted to our geometry.
In spite of the presence of rotation, the photon bubble instability survives
unscathed in this limit.

Finally, if we set the equilibrium magnetic field equal to zero, the
equilibrium density to be constant, and the gas pressure to be negligible,
then after cancelling two resulting zero modes, the dispersion relation reduces 
to fifth order:
\begin{equation}
3\omega^5+i{k^2c\over\kappa_{\rm es}\rho}\omega^4-\left(3\kappa^2+
{4k^2E\over3\rho}\right)\omega^3-i{k^2c\kappa^2\over\kappa_{\rm es}\rho}\omega^2
+\left({4k_z^2\kappa^2E\over3\rho}-3k_R^2g^2\right)\omega+
{k_zk_R^2gc\kappa^2\over\kappa_{\rm es}\rho}=0.
\label{pkdisp}
\end{equation}
This equation resembles the dispersion relation presented by Pietrini \&
Krolik (2000) when the dissipation terms that they included are neglected.
Their dispersion relation still has several additional terms compared to
equation (\ref{pkdisp}), however.  Each of these terms are small and arise
from the left hand side of the full radiation momentum equation
(\ref{dfdt}), which we neglected.

Despite the complexity of the general dispersion relation, the MRI turns
out to be a robust instability.  An analytic indication that this is so can
be found by just setting $\omega=0$ in equation (\ref{stratdisp}), which
leads to the following equation:
\begin{equation}
k_z^2\vaz^2\left(k^2\vaz^2+R{d\Omega^2\over dR}\right)
\left[i{k_zk_R^2F_zk_{\rm diff}^2\over3\rho k^2\kappa}+
\left(g-{k_z^2F_zk_{\rm diff}^2\over3\rho k^2\kappa}\right){1\over\rho}
{\partial\rho\over\partial z}-k_z^2\cgas^2\right]-
k_R^2k_z^2\vaz^2\cgas^2N_{\rm g}^2=0.
\end{equation}
If the gas Brunt-V\"ais\"al\"a frequency $N_{\rm g}=0$,
this equation shows that there will exist a zero frequency root if
$k<k_{\rm crit}\equiv[-R(d\Omega^2/dR)]^{1/2}/\vaz$, the usual
critical wavenumber for
the onset of the MRI.  Even if $N_{\rm g}\ne0$, there will be a zero
frequency root for $k_R=0$ and $k^2\vaz^2=-R(d\Omega^2/dR)$.  While not
a rigorous proof, these facts suggest that the critical wavenumber for
the onset of the MRI is unchanged by the effects of radiation.   Hence
the instability criterion for the MRI is still that the angular velocity
decreases outward in magnitude.   These expectations are born out by
our numerical calculations.   Radiative diffusion does not affect the
stability criterion, although it can dramatically affect the growth rates,
as discussed in section 3.

\subsection{Photon Bubble Modes and Other MHD Wave Instabilities}

In the high wavenumber limit $k\rightarrow\infty$, the dispersion
relation can be factored to give
\begin{equation}
\left(\omega+i{k^2c\over3\kappa_{\rm es}\rho}\right)(\omega^2-k_z^2\vaz^2)
[\omega^4-\omega^2k^2(\cgas^2+\vaz^2+\vaphi^2)+k_z^2\vaz^2k^2\cgas^2]=0.
\end{equation}
The first factor is a strongly damped radiation diffusion mode, the second
describes Alfv\'en waves, and the third corresponds to the fast and slow
magnetosonic waves.  Rotation is irrelevant at these high wavenumbers.
We have lost the radiation effects on the fast and slow modes because in
this limit the wavelength is much less than the diffusion length scale.

It is worthwhile examining the behavior of the MHD wave
modes to higher order in $k^{-1}$, because they all turn out to be unstable
for sufficiently high vertical photon fluxes.
If, as we have been generally assuming, the radiation and magnetic energy
densities in the plasma are much larger
than the gas thermal energy density, then the fast and slow modes
have frequencies given by
\begin{equation}
\omega\simeq\pm k(\vaz^2+\vaphi^2)^{1/2}-i{3\kappa_{\rm es}\rho\over2c
(\vaz^2+\vaphi^2)}\left[\crad^2\left({k_R^2\over k^2}\vaz^2+\vaphi^2\right)
\mp{k_zk_R^2F_z\vaz^2\over3k^3\rho(\vaz^2+\vaphi^2)^{1/2}}\right]
\label{fastdisp}
\end{equation}
and
\begin{equation}
\omega\simeq\pm k_z\cgas\left({B_z\over B}\right)\mp i{\kappa_{\rm es}k_R^2 F_z
\over2 c\cgas k^2}\left({B_z\over B}\right),
\label{slowdisp}
\end{equation}
respectively.  The lowest order correction to the Alfv\'en mode is of order
$k^{-1}$, and this correction is real, implying that any instability or
damping is at still higher order.  As we shall see, instabilities in the
Alfv\'en waves peak at finite wavenumbers.

The first imaginary term in equation (\ref{fastdisp}) represents damping
of the fast modes by radiative diffusion.  This term agrees with the
optically thin damping rate derived by Agol \& Krolik (1998), although the
waves need not be optically thin for this rate to apply.  The last term
in equation (\ref{fastdisp}) shows that a sufficiently large vertical
radiative flux can overcome the damping and destabilize the fast waves.
The instability criterion can be obtained by comparing the last two terms
of equation (\ref{fastdisp}):
\begin{equation}
F_z>3\rho\crad^2\left(1+{k^2\vaphi^2\over k_R^2\vaz^2}\right){k\over k_z}
(\vaz^2+\vaphi^2)^{1/2}.
\label{fastcrit}
\end{equation}
Crudely, if the speed at which photons are diffusing upward in the equilibrium
exceeds the Alfv\'en speed, then fast modes propagating {\it upward} at some
nonzero angle to the vertical will be unstable.  For large photon fluxes,
the growth rate of this instability is
\begin{equation}
{\rm Im}(\omega)\simeq{g\over2 v_A}\left({k_zk_R^2\over k^3}\right)
\left({B_z\over B}\right)^2.
\end{equation}
The instability criterion is most easily satisfied, and the largest growth
rates are achieved, for vertical magnetic fields.  Azimuthal fields tend
to be stabilizing.

Equation (\ref{slowdisp}) reveals that slow modes propagating {\it downward}
at some nonzero angle to the vertical are also unstable at high photon fluxes.
These modes have the fastest growth rates, and at lower values of $k$ become
the photon bubble modes described by Gammie's (1998) dispersion relation
(\ref{pbdisp}).
Adopting Gammie's (1998) photon bubble ansatz of $\omega\sim k^{1/2}$ and
retaining only the leading order terms as $k\rightarrow\infty$ in the general
dispersion relation, we find
\begin{equation}
\omega^2=\left(k_z^2\cgas^2-i{k_zk_R^2\kappa_{\rm es}F_z\over ck^2}
-i\omega{3\kappa_{\rm es}\rho\crad^2k_z^2\over ck^2}\right)
\left({B_z\over B}\right)^2.
\end{equation}
The middle term gives the photon bubble dispersion relation (\ref{pbdisp})
when use is made of the hydrostatic equilibrium condition (\ref{hydrostat})
in the radiation dominated limit.  The other terms compete with the photon
bubble term at low and high wavenumbers, leading us to expect that the
photon bubble dispersion relation (\ref{pbdisp}) will only hold in
the range of wavenumbers given by
\begin{equation}
{9\rho^2\crad^2\kappa_{\rm es}^2\over c^2g}\lta k\lta{g\over\cgas^2}.
\label{pbcond}
\end{equation}
A condition equivalent to this was also noted by Gammie (1998).  When the
last inequality breaks down at high wavenumbers, the slow mode dispersion
relation is better described by equation (\ref{slowdisp}).

Figure 3 presents a numerical solution to the full dispersion relation
(\ref{stratdisp}) which illustrates these high wavenumber instabilities
and compares their growth rates to the analytic expressions above.
In addition to
the fast and slow modes, the Alfv\'en modes also appear to go unstable
for large photon fluxes.  In contrast to the fast and slow modes, the
Alfv\'en wave growth rates tend to zero as $k\rightarrow\infty$.  Rotation
directly affects the unstable growth rates of the Alfv\'en modes,
and also affects the fast and slow modes at low wavenumbers.  We describe this,
and the coupling to the MRI, in the next subsections.

\subsection{Zero Azimuthal Field Case}

We now return to low wavenumbers where rotation is important.  We first
consider the case where the equilibrium field is vertical ($B_\phi=0$).
The general dispersion relation (\ref{stratdisp}) then reduces to
\begin{eqnarray}
& &D_{\rm ms}D_{\rm BH}-{k_R^2\over k_z^2}\tilde\omega^2\omega^4
+\left(D_{\rm BH}-{k_R^2\omega^2\tilde\omega^2\over k_z^2}\right)
\left[-{k_zk_R^2F_z\over3\omega_{\rm diff}\rho}+
\left(g-i{k_z^2F_z\over3\omega_{\rm diff}\rho}\right){1\over\rho}
{\partial\rho\over\partial z}\right]\nonumber\\
& &+k_R^2\tilde\omega^2\left(\cgas^2N_{\rm g}^2+{\omega\over\omega_{\rm diff}}
\crad^2N_{\rm r}^2\right)
+{k_R^2k_z\tilde\omega^2F_z\omega^2\over3\omega_{\rm diff}\rho}=0.
\label{vphizero}
\end{eqnarray}
Note that $D_{\rm BH}=0$ is a solution of this equation when $k_R=0$, so we
immediately conclude that in the absence of an azimuthal field, the vertical
MRI modes are completely unaffected by compressibility, stratification, and
radiation.  This is not surprising, as the vertical MRI modes involve
horizontal motions only, and therefore do not involve any excess forces
from the vertical radiative flux.

Equation (\ref{vphizero}) is complicated, with many parameters.  For
numerical computations, we restrict consideration to cases where the gas pressure
is negligible, so that $\cgas$ and $N_{\rm g}$ are both zero.  In
addition, we neglect convection and couplings to gravity modes by considering
only isentropic equilibria, so that $N_{\rm r}=0$.  Under these assumptions,
we may use equation (\ref{hydrostat}) to write the acceleration due to gravity
and the equilibrium vertical
density gradient in terms of the photon
flux, viz.
\begin{equation}
g={k_{\rm diff}^2\over3\kappa\rho}F_z\,\,\,\,{\rm and}\,\,\,\,{1\over\rho}
{\partial\rho\over\partial z}=-{k_{\rm diff}^2\over3\kappa\rho\crad^2}F_z.
\end{equation}
Hence the only new parameter is $F_z$.

Figure 4 presents numerical calculations of the unstable mode growth
rates in the diffusive regime at low wavenumber, for various values
of photon flux.  The introduction of stratification causes the slow (photon
bubble) and Alfv\'en wave instabilities to appear and have increasing
growth rate as $F_z$ gets larger.  The fast wave instability appears
only after a critical threshold in $F_z$ is reached [cf. Fig. 4(b)].
This is quantitatively consistent with our instability criterion
(\ref{fastcrit}), which predicts that the fast wave will be unstable
for $F_z/(3\rho\vaz^3)>125$ for the parameters chosen in Fig. 4.
Fig. 4(c) should be compared with Fig. 3 which has similar radiation
parameters.  Note that for the value of $k_{\rm diff}$ chosen in Fig. 4,
the MRI converts to Alfv\'en waves at $k_{\rm crit}$ [cf. Fig. 2(b)], and
indeed the unstable Alfv\'en waves show up for $k>k_{\rm crit}$.  The
MRI itself appears to be largely unaffected by stratification, except
that the maximum growth rate increases with increasing $F_z$.  It
is unclear what would happen at higher fluxes because we are already
close to the WKB limit (\ref{wkbcond}) at these high fluxes.  Given
that the vertical ($k_R=0$) MRI modes are unaffected by flux, they
are not always the fastest growing modes in the presence of nonzero
flux.  Modes with a slight nonzero $k_R$ in fact grow faster for
the equilibrium parameters shown in Fig. 4(c).  The fact that $F_z$ can
affect the MRI growth rates for $k_R\ne0$, albeit modestly, suggests
that geometrically thick accretion flows with significant radial flux
in the equilibrium could have a more marked effect on the MRI than the
geometrically thin configurations considered in this paper.

The photon bubble mode is always the most rapidly growing of the three
MHD wave instabilities that exist in stratified media.  It appears
to survive for $k<k_{\rm crit}$, albeit at a smaller growth
rate than expected from extrapolation of the high $k$ behavior.
Rotation therefore acts to mildly suppress, and does not kill,
this instability when $B_\phi=0$.

\subsection{Nonzero Azimuthal Field}

We finally turn to the full dispersion relation (\ref{stratdisp}) with
nonzero azimuthal field, but continuing to neglect gas pressure
and vertical entropy gradients.  Figure 5 presents results for the
low wavenumber instability growth rates corresponding the the $B_\phi=0$
case shown in Fig. 4(b).  As we stated above, the MRI is a robust
instability which even in the presence of a nonzero azimuthal field
and radiation, still exists for $k<k_{\rm crit}$.  However, as expected
from our results in section 3, the nonzero azimuthal field produces
a marked reduction in the overall growth rate.  The maximum MRI growth
rates in Fig. 5 are comparable to the $k_R=0$ growth rates shown
in Fig. 1 at the chosen diffusion wavenumber $k_{\rm diff}\vaz/\Omega=0.1$.
The vertical radiative flux makes little difference to the MRI compared
to the nonstratified case, just as we found in the previous subsection.

The azimuthal field component has a noticeable effect on the other instabilities.
First, the fast and Alfv\'en waves are quickly stabilized.  This is consistent
with our findings in section 4.1, where we noted that azimuthal fields
would stabilize the fast waves.  For the parameters chosen in Fig. 5,
equation (\ref{fastcrit}) predicts fast wave stability for
$B_\phi\gta0.4B_z$.

The photon bubble growth rate also decreases with increasing azimuthal
field, in quantitative agreement with equation (\ref{pbdisp}).  At
high values of $B_\phi/B_z$, this mode appears to split into a
high wavenumber part (the true photon bubble), and a low wavenumber
part [cf. Fig. 5(b)].  The latter dominates the MRI growth rate
except at the lowest wavenumbers, at least for the parameters chosen
in Fig. 5.  At high wavenumbers, this mode evolves into the Alfv\'en mode.

\section{Astrophysical Relevance}

It turns out that diffusion is actually very important in standard
$\alpha$-viscosity models of accretion disks around black holes and
neutron stars (Shakura \& Sunyaev 1973).  Neglecting relativistic
correction factors, one can show that in the radiation pressure
dominated inner zone,
\begin{equation}
k_{\rm diff}H\simeq2\alpha^{-1/2}\,\,{\rm radians}.
\end{equation}
Surprisingly, this relation is independent of accretion rate, radius, and
central object mass.  Combining this with the WKB condition for the
MRI, $k\sim\Omega/\vaz\gg H^{-1}$, we find
\begin{equation}
{k_{\rm diff}\vaz\over\Omega}\ll\alpha^{-1/2}.
\label{eqkdiff}
\end{equation}
Depending on the assumed strength of the initial magnetic field, realistic
accretion disk models ($\alpha\sim0.01-0.1$) will often be in the regime where
photon diffusion can significantly slow the growth of the MRI in the
presence of azimuthal field components (cf. Fig. 1).  Our analysis shows
that in this case the growth rate would be reduced to
$\sim\Omega(\cgas/\vaphi)$.  In the radiation pressure dominated inner
zone, one can conceive of $\vaphi$ being somewhat less than $\crad$.
Neglecting relativistic correction factors once again,
\begin{equation}
{\cgas\over\crad}\simeq4\times10^{-3}\alpha^{-1/8}\eta\left({M\over\msun}
\right)^{-1/8}\left({L\over\ledd}\right)\left({R\over r_{\rm g}}
\right)^{21/16},
\end{equation}
where $M$ is the mass of the central object.  The MRI growth rate can
therefore become quite sluggish if the azimuthal field energy density
exceeds the thermal energy density in the gas by an amount comparable to
the radiation/gas thermal energy density ratio.  We stress however that
even in this regime, the MRI still exists as a dynamical instability,
and will still presumably grow eventually to nonlinear amplitude.  It
might be interesting to see if the resulting turbulence depends on how
the initial magnetic field energy density compares to the gas and radiation
thermal energy densities, but it is difficult to predict any results
from linear theory alone.

Note that all these conclusions refer to a fictitious, unstable initial
state which cannot exist in nature.  The equations above, in particular
that for the diffusion wavenumber (\ref{eqkdiff}), depend on $\alpha$,
which is presumably a consequence of the MRI.
It is therefore difficult to determine how the MRI will be affected
when it is responsible for the ambient conditions in the disk in the
first place.

This fact is even more important when we consider the stratified calculations
that we presented in section 4.  Multiple instabilities are simultaneously
present in this case, and the modes with the fastest growth rates
presumably dominate the nonlinear evolution.  One would at first conclude
that the photon bubble/slow mode at high wavenumbers therefore controls
the dynamics.  From equation (\ref{slowdisp}), the growth rate is at
most
\begin{equation}
{\rm Im}(\omega)\simeq10^2\Omega\alpha^{1/8}\eta^{-1}
\left({L\over\ledd}\right)\left({M\over\msun}\right)^{1/8}
\left({R\over r_{\rm g}}\right)^{-21/16},
\label{pbhighk}
\end{equation}
substantially faster than the orbital frequency $\Omega$.

However, it is important to recognize the
different sources of free energy which are driving these instabilities.
The MRI feeds off the differential rotation in the flow, transporting
angular momentum outward which ultimately allows material to accrete
and release gravitational potential energy.  We therefore view this
instability as being fundamental to the dynamics of the accretion disk.
Dissipation of the gravitational energy released and radiative
transport processes set up vertical gradients which drive the photon
bubble and other MHD wave instabilities.  Convection will also presumably
occur if the resulting entropy gradient has the right sign (Pietrini
\& Krolik 2000).  It is the vertical gradients which provide
the source of free energy for these extra instabilities, but these gradients
are fundamentally determined by the action of the MRI.

We speculate that on the largest vertical scales, where diffusion is
negligible on an orbital time, the MRI will always operate as usual,
launching perhaps a turbulent cascade to smaller scales.  How this
energy is ultimately dissipated is still unclear.  If there is sufficient
vertical flux, it may not be as simple as radiative damping of compressive
MHD waves (Agol \& Krolik 1998), as those waves are themselves unstable
for some propagation directions.  Given the high growth rates at
short wavelengths from equation (\ref{pbhighk}) above, it seems likely to
us that photon bubbles will play an important role in thermal transport
on small scales.

Concrete answers to the question of how radiative diffusion and radiative flux
really affect the turbulent angular momentum and heat transport in the
disk will require investigation by numerical simulations.  One must be
careful to choose the fictitious unstable initial state with care when
pursuing such simulations, keeping in mind that accretion
power, and therefore the MRI, is the driver for all this complex dynamics.

\section{Conclusions}

We may summarize our conclusions as follows.
In the absence of an azimuthal field, the vertical ($k_R=0$) MRI modes
have the standard, incompressible behavior elucidated by Balbus \& Hawley
(1991).  If the azimuthal field component dominates the vertical field
component and the Alfv\'en speed is larger than the effective sound speed
[cf. eq. (\ref{vaphicond})], then the growth rate of the MRI is reduced
to $\sim\Omega(\cgas/\vaphi)$, well below the usual rapid growth rate of
order the orbital frequency $\Omega$.  However, the instability still
persists.  Large vertical photon fluxes in the equilibrium do not
significantly affect the most rapidly growing MRI modes.

At short wavelengths where rotation is unimportant, both the fast and
slow MHD waves are destabilized by sufficiently high vertical photon flux
in the equilibrium.  These instabilities require that the waves propagate
at some nonzero angle to the vertical, and affects upward propagating
fast waves and downward propagating slow waves.  The slow waves appear to
always have the larger growth rates.  At smaller wavenumbers these are
well-described by Gammie's (1998) photon bubble dispersion
relation (\ref{pbdisp}), but at larger wavenumbers they asymptote to a maximum
value given by equation (\ref{slowdisp}).  At small wavenumbers where
rotation is important ($k<k_{\rm crit}$),
the photon bubble growth rate is somewhat suppressed by rotation, at least at
modest fluxes.  Nonzero azimuthal fields stabilize the fast wave instability
and reduce the photon bubble growth rate.  In addition to the fast and
slow wave instabilities, the vertical radiative flux also drives unstable
modes at intermediate wavenumber which are related to Alfv\'en
waves at high wavenumber.

The ambient conditions in the radiation pressure dominated inner region
of standard alpha disk models are generally in a regime where radiative
diffusion can slow the growth of the MRI.  In addition, photon bubble
modes are expected at short wavelengths with growth rates that can
significantly exceed the orbital frequency.  It is not clear how these
instabilities will evolve into the nonlinear, turbulent regime, but
we speculate that the MRI will dominate the largest scales and that
photon bubbles (and perhaps unstable fast and Alfv\'en waves) may affect
both the dissipation and thermal transport on smaller scales.
A complete understanding
of the turbulent angular momentum and heat transport will
require numerical simulations.

Finally, we stress that we have concentrated here on geometrically
thin accretion disk configurations where the photon flux is primarily
in the vertical direction and the rotation curve is near-Keplerian.
Geometrically thick or quasi-spherical, radiation pressure supported
flows will have radiation pressure gradient components in the radial
direction and may therefore exhibit more interesting dynamical coupling
between the radiative flux and the MRI.

\acknowledgments{We acknowledge useful conversations with Eric
Agol, Yuri Levin, and Jim Stone.  This work was supported
by NSF grants AST-9970827 and PHY-9907949, and NASA grant NAG5-7075.}
 
\appendix

\section{Appendix A: The Zero Frequency Mode of the General Dispersion Relation}

Despite the fact that the perturbation equations (\ref{dcont})-(\ref{dbphi})
and (\ref{divdb})
have eight separate $\omega$-dependent terms, we only end up with
a dispersion relation that is seventh order.  The reason is that there
is a zero frequency mode, which we discuss in some detail in this appendix.

Setting $\omega=0$ in the perturbation equations, the radial induction equation
(\ref{dbr}) immediately implies that $\delta v_R$ must vanish.  The continuity
equation (\ref{dcont}) and azimuthal momentum equation (\ref{dvphi}) then
imply that $\delta v_z$ and $\delta B_\phi$ must also vanish.  The gas
energy equation (\ref{dengas}) is then trivially satisfied, which is the
mathematical reason why a zero frequency mode is a possible solution of
the perturbation equations.

Excluding the trivial azimuthal flux equation (\ref{dfphi}), the remaining
seven equations may be written
\begin{equation}
-2\Omega\rho\delta v_{\phi}
+ik_{R}\left(\delta p+{1\over3}\delta E\right)
+i{k_{R}\over 4\pi}B_{z}\delta B_{z}-i{k_{z}\over 4\pi}B_{z}\delta B_{R}=0,
\end{equation}
\begin{equation}
ik_{z}\left(\delta p+{1\over3}\delta E\right)+g\delta\rho=0,
\end{equation} 
\begin{equation}
{\bf k}\cdot\delta {\bf F}=0,
\end{equation}
\begin{equation}
\delta F_R+{ic\over 3\kappa_{es}\rho}k_R \delta E=0,
\end{equation}
\begin{equation}
\delta F_z+F_z{\delta\rho\over\rho}+{ic\over 3\kappa_{es}\rho}k_z \delta E=0,
\end{equation}
\begin{equation}
-R{d\Omega\over dR}\delta B_{R}-ik_{z}B_{z}\delta v_{\phi}=0,
\label{zshear}
\end{equation}
and
\begin{equation}
k_r\delta B_R + k_z\delta B_z = 0.
\end{equation}
These seven equations have eight unknowns, so that any one of the nonzero
perturbation amplitudes determine the other seven for this zero frequency
mode.

It is perhaps surprising that this mode has such a complicated eigenfunction,
and it is therefore worthwhile understanding its physical meaning.
Equation (\ref{zshear})
is particularly interesting.  This arose from the azimuthal component of
the induction equation which describes, for example, how a radial magnetic
field component is sheared out by the differential rotation to produce a
time-dependent azimuthal field component.  In this case, however, this
is exactly compensated by an azimuthal velocity perturbation which creates
azimuthal field in the opposite direction out of vertical field.

Further insight into the physical nature of this mode may be obtained
by taking various limits.  In the limit $k_z\rightarrow0$, we have
$\delta B_R=\delta\rho=\delta E=\delta{\bf F}=0$, and the only nontrivial
condition comes from the radial momentum equation:
\begin{equation}
-2\Omega\rho\delta v_{\phi}
+ik_{R}\delta p
+i{k_{R}\over 4\pi}B_{z}\delta B_{z}=0.
\end{equation}
In other words, any random field of perturbations of $\delta v_\phi$ can
be balanced by radial gradients of gas and/or magnetic pressure to produce
a slightly different equilibrium.

In the opposite limit $k_R\rightarrow0$, we have
$\delta v_\phi=\delta B_z=\delta B_R=\delta F_z=0$, and
\begin{equation}
ik_{z}\left(\delta p+{1\over3}\delta E\right)+g\delta\rho=0,
\end{equation} 
\begin{equation}
\delta F_R+{ic\over 3\kappa_{es}\rho}k_R \delta E=0,
\end{equation}
\begin{equation}
F_z{\delta\rho\over\rho}+{ic\over 3\kappa_{es}\rho}k_z \delta E=0.
\end{equation}
In this limit, a random field of density perturbations is balanced by
vertical gradients in gas and radiation pressure, again producing a
slightly different equilibrium.

Finally, we note that if we neglect the vertical stratification of the
equilibrium, then this zero mode has all perturbations vanishing except
for the density.  In this case a random field of density perturbations
will merely be advected along with the equilibrium flow without producing
additional net accelerations.

\begin{figure}
\figurenum{1}
\plotone{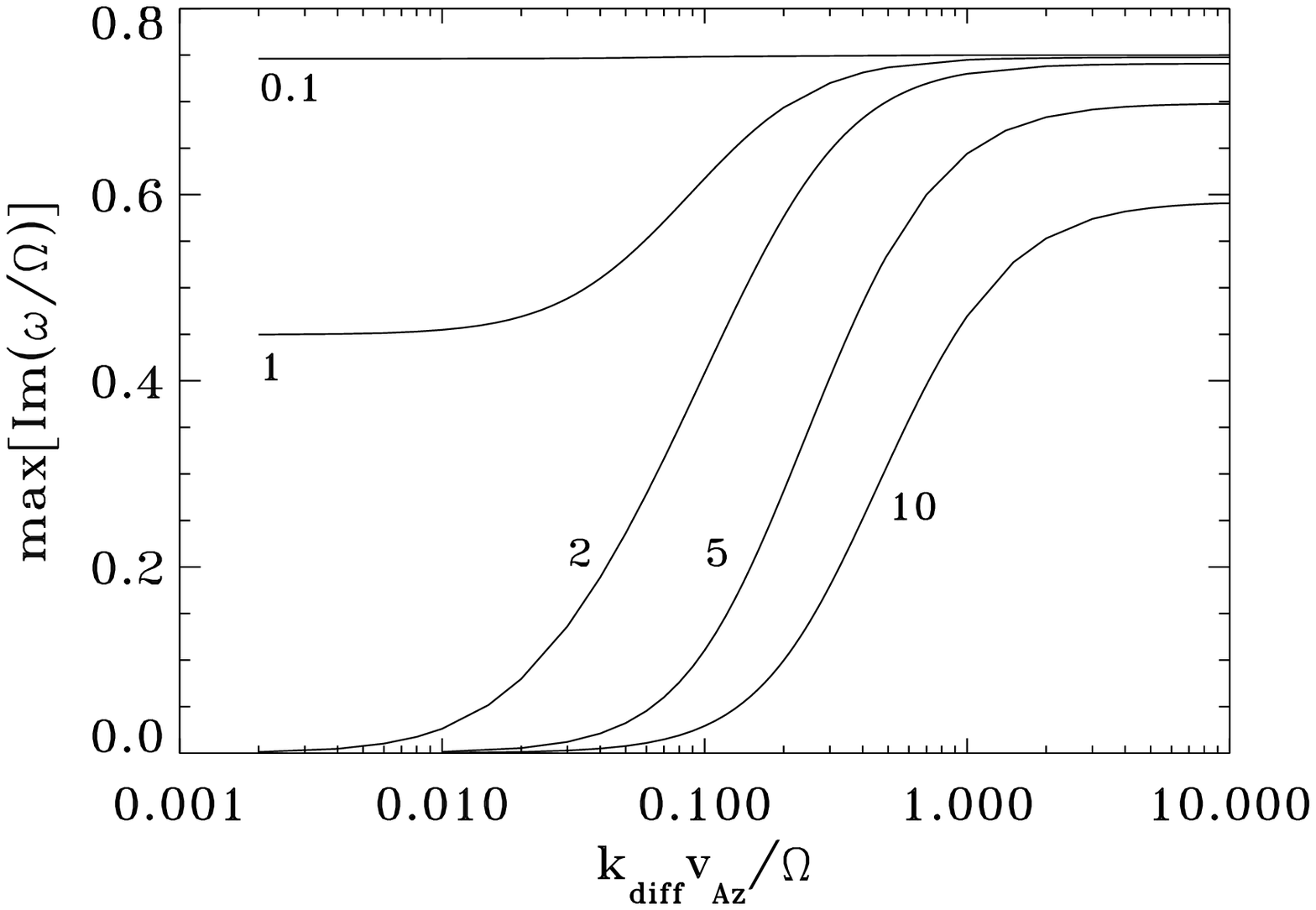}
\vskip 0.1truein
\caption{Maximum growth rate of the MRI in a medium with no stratification,
as a function of azimuthal field strength and diffusion wavenumber.  We have
adopted a Keplerian rotation curve ($\kappa=\Omega$) for these calculations,
and have set $\cgas=0$, $\crad/\vaz=10$, and $k_R=0$.  Each curve corresponds
to different values of $\vaphi/\vaz$, which are labelled.  Note the substantial
reduction in growth rate at modest azimuthal field strengths and low diffusion
wavenumber.}
\end{figure}

\vfill\eject

\begin{figure}
\figurenum{2}
\epsscale{0.7}
\plotone{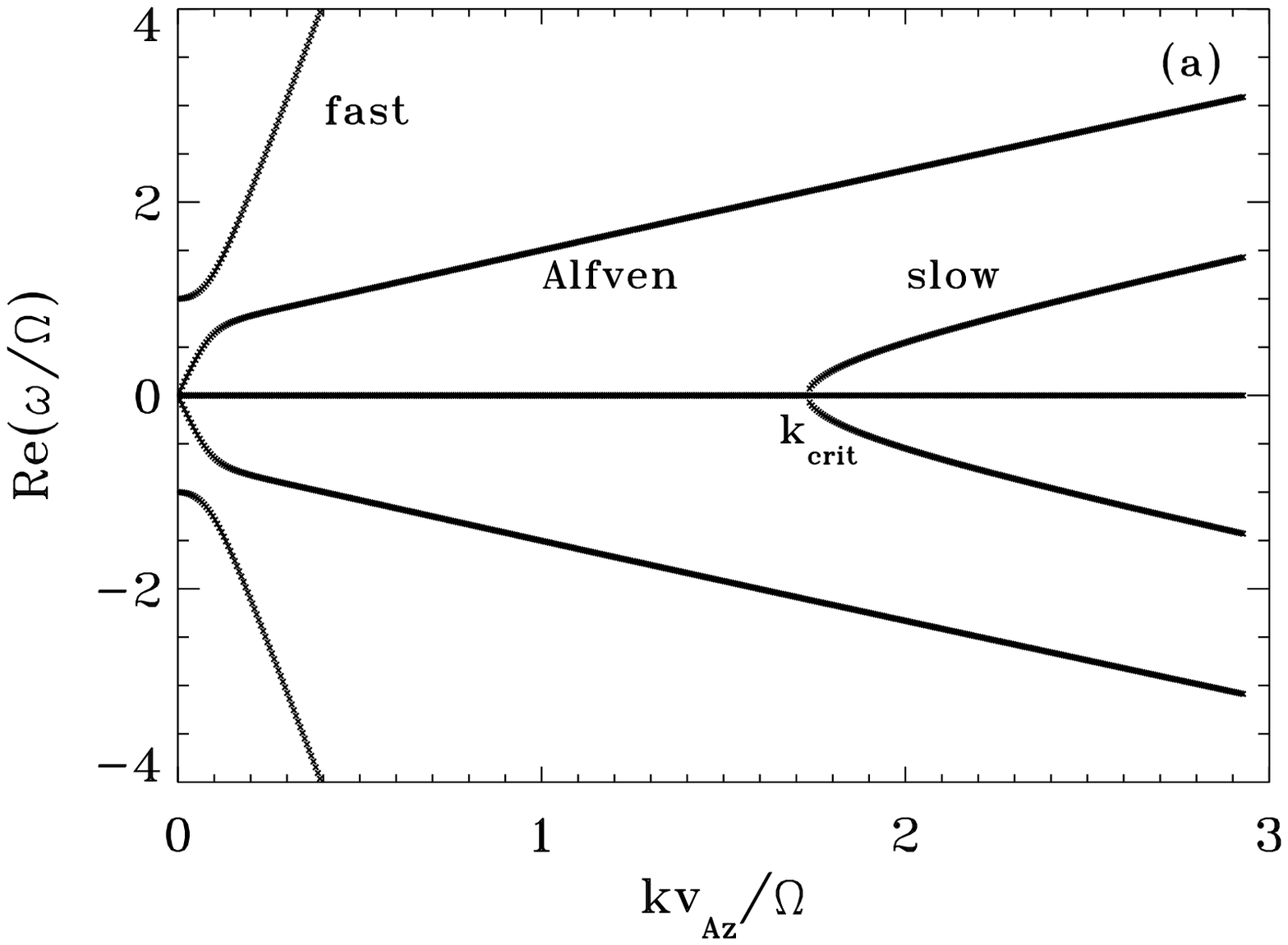}
\end{figure}

\begin{figure}
\figurenum{2}
\epsscale{0.7}
\plotone{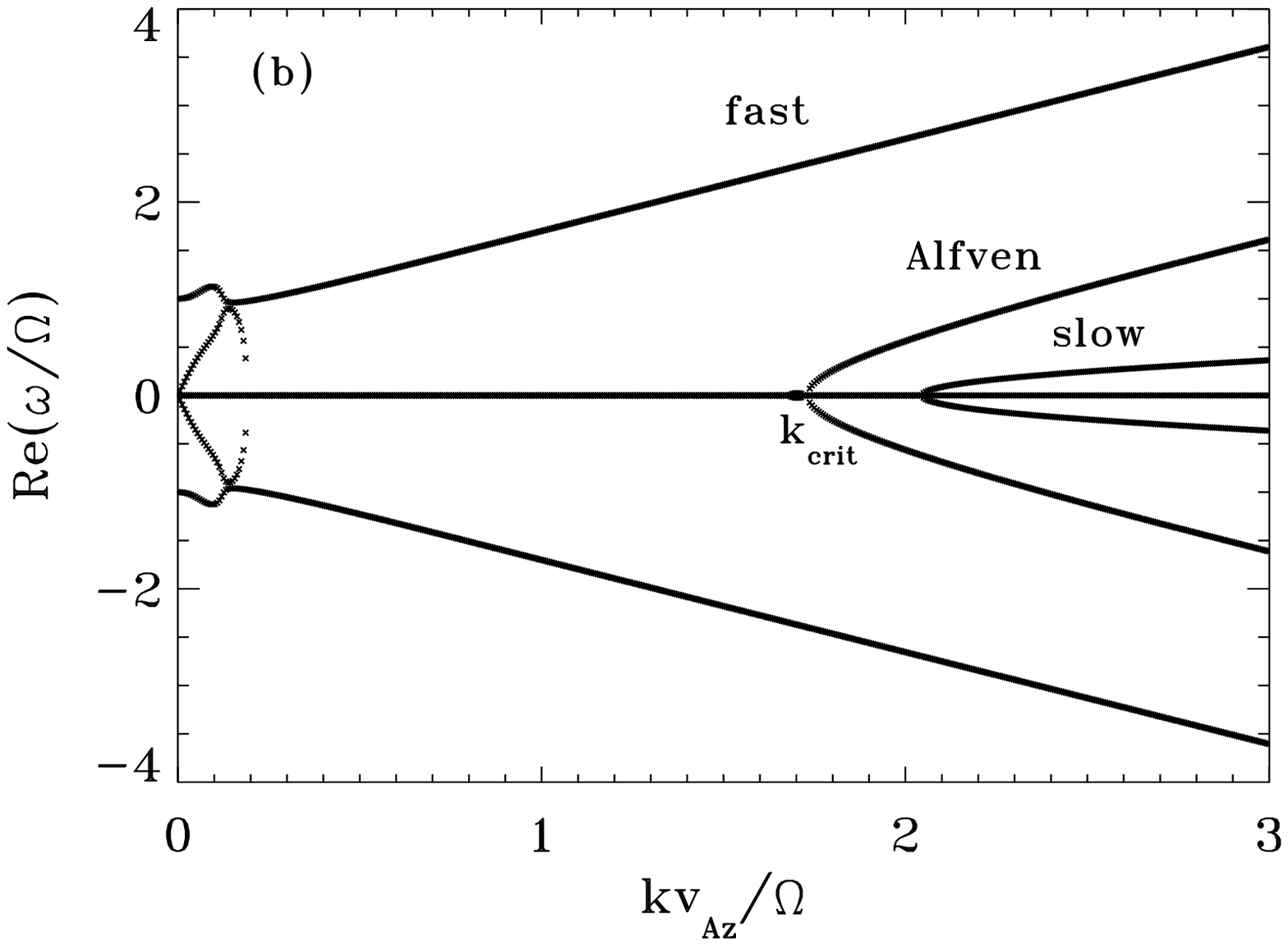}
\end{figure}

\begin{figure}
\figurenum{2}
\epsscale{0.7}
\plotone{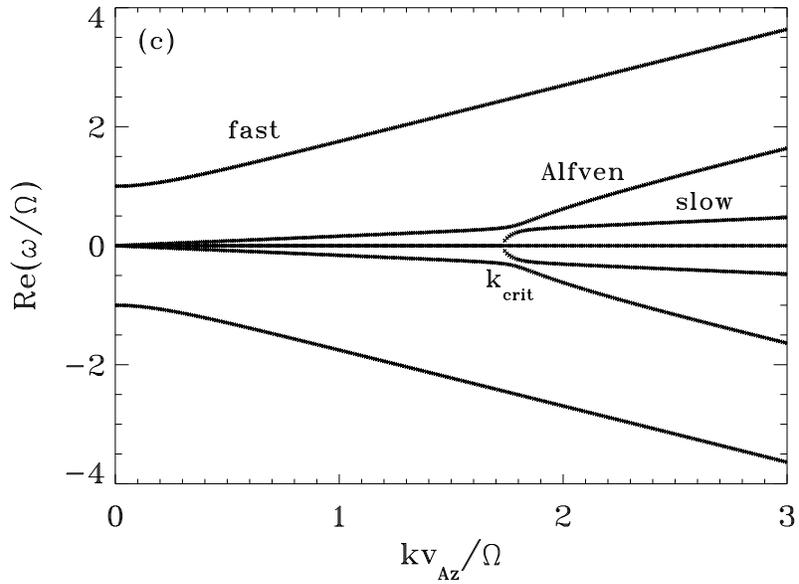}
\vskip 0.1truein
\caption{Real part of the mode frequencies in a medium with no stratification
and a vertical magnetic field ($B_\phi=0$), as a function of total wavenumber.
We have assumed a Keplerian rotation curve and $k_z/k=0.8$, $\crad/\vaz=10$,
and $\cgas/\vaz=0.2$.  Each figure corresponds to a different value of
$k_{\rm diff}\vaz/\kappa$: (a) 10, (b) 0.1, and (c) 0.01.  The critical
wavenumber $k_{\rm crit}\equiv[-R(d\Omega^2/dR)]^{1/2}/\vaz$ above which
the MRI is stabilized is indicated in each figure.}
\end{figure}

\vfill\eject

\begin{figure}
\figurenum{3}
\epsscale{1}
\plotone{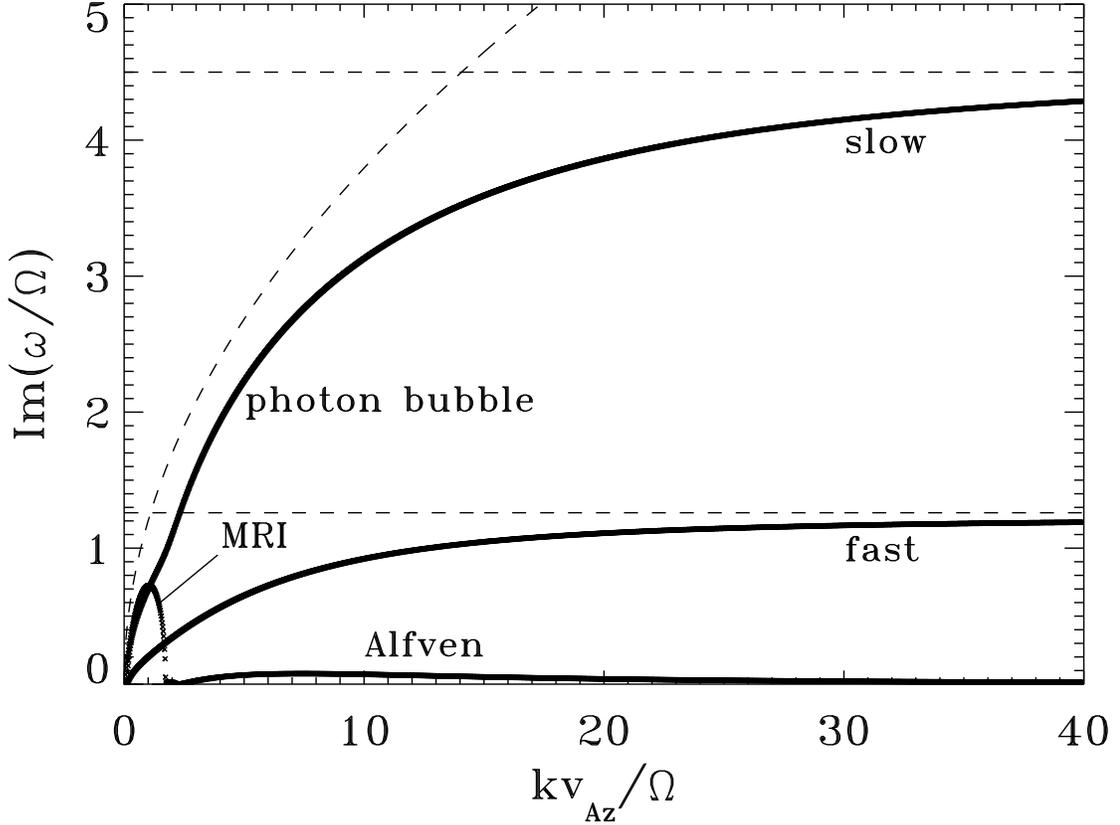}
\vskip 0.1truein
\caption{Imaginary part of the mode frequencies in a stratified medium with
vertical magnetic field ($B_\phi=0$) as a function of total wavenumber.
The rotation curve is Keplerian and  $k_z/k=0.8$, $\crad/\vaz=10$,
$\cgas/\vaz=0.4$, $k_{\rm diff}\vaz/\kappa=0.1$, and $F_z/(3\rho\vaz^3)=1000$.
The dashed curves show the analytic approximations (\ref{pbdisp}),
(\ref{fastdisp}), and (\ref{slowdisp}).
[We have neglected the Brunt-V\"ais\"al\"a terms in the dispersion relation
(\ref{stratdisp}) and also set $g/(\kappa\vaz)=10$ and
$(g/\kappa^2)d\ln\rho/dz=-1$.  However, these last two parameters do not
significantly affect the results and could just as well have been set to
zero.  This is consistent with the WKB approximation.]}
\end{figure}

\vfill\eject

\begin{figure}
\figurenum{4}
\epsscale{0.7}
\plotone{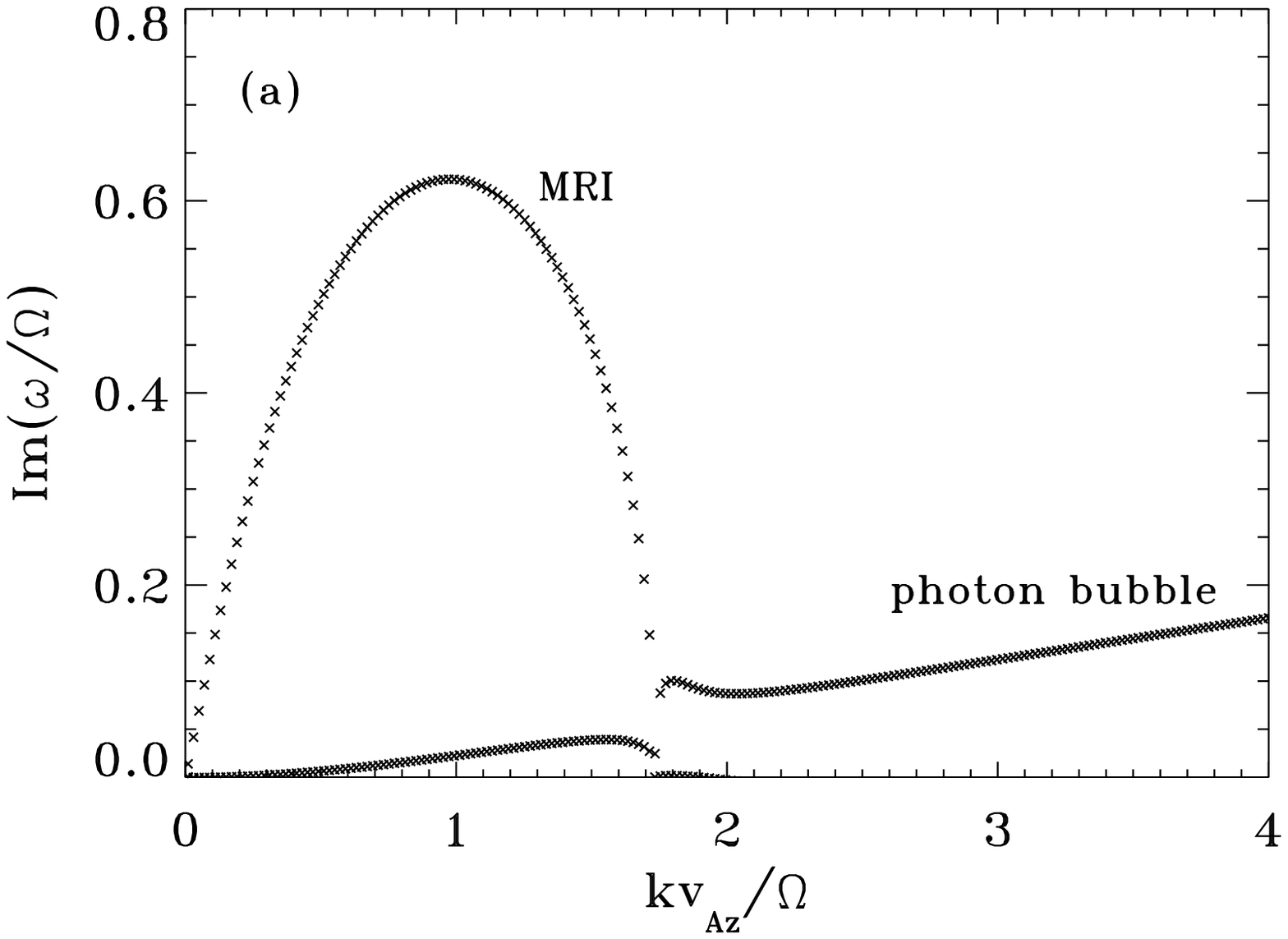}
\end{figure}

\begin{figure}   
\figurenum{4}
\epsscale{0.7}
\plotone{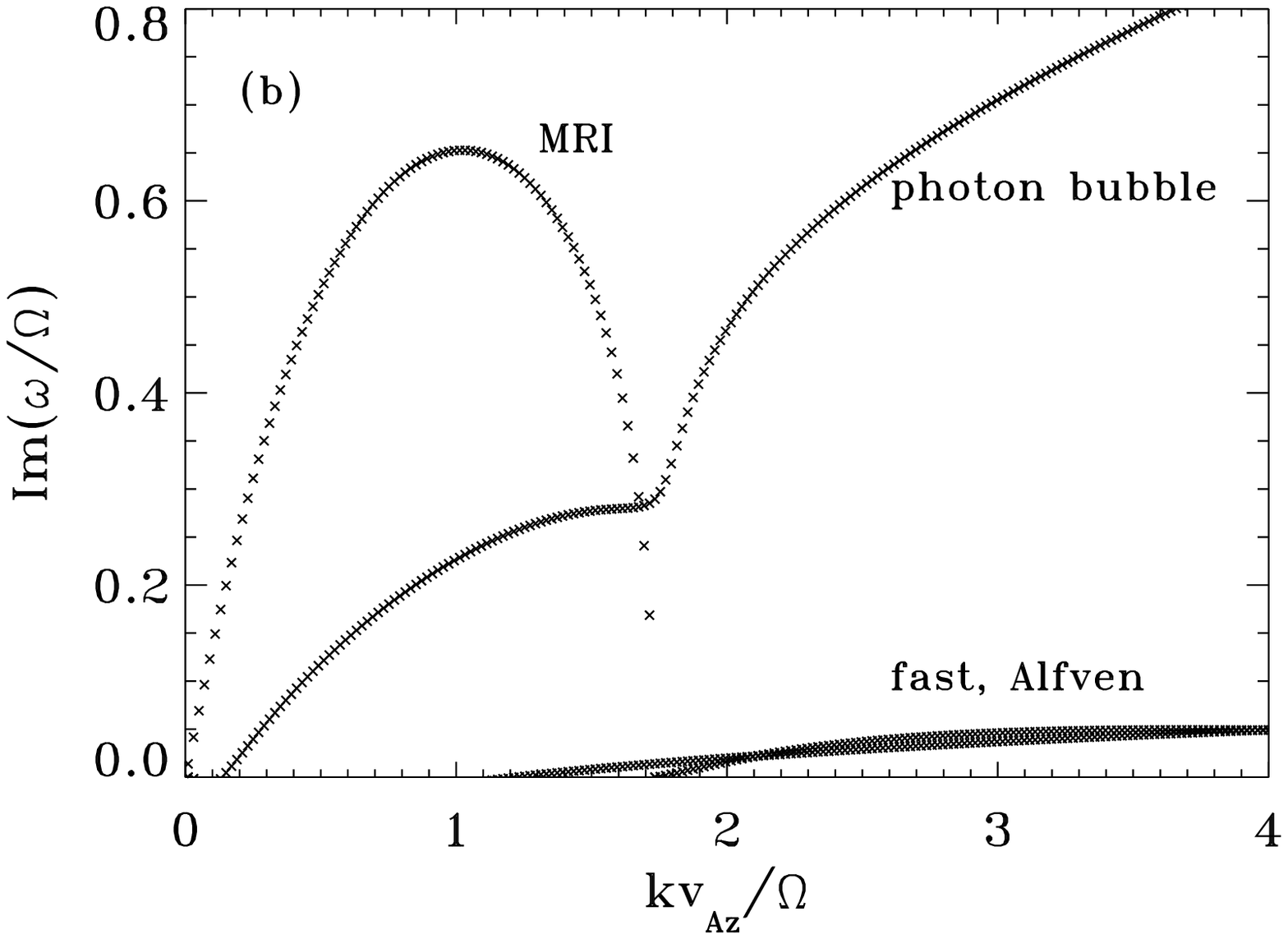}
\end{figure} 

\begin{figure}   
\figurenum{4}
\epsscale{0.7}
\plotone{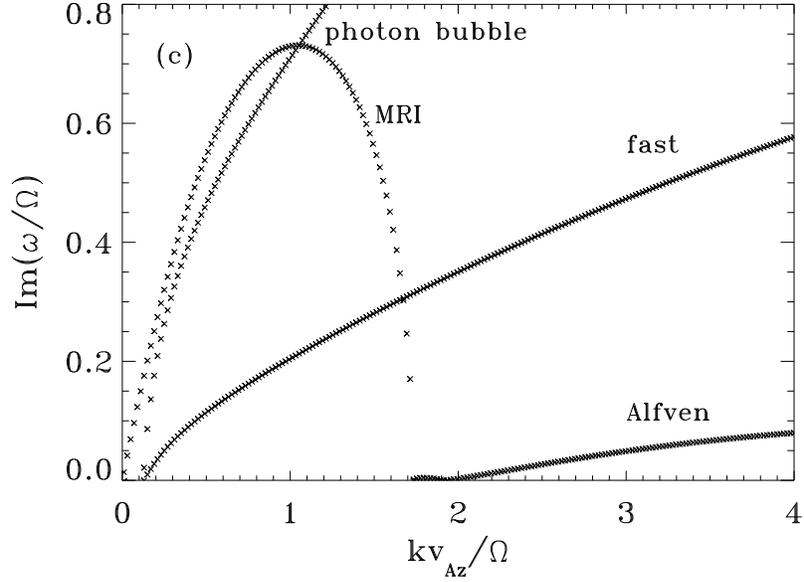}
\vskip0.1truein
\caption{Imaginary part of the mode frequencies in a stratified medium with
vertical magnetic field ($B_\phi=0$) as a function of total wavenumber.
The rotation curve is Keplerian and  $k_z/k=0.8$, $\crad/\vaz=10$,
$k_{\rm diff}\vaz/\kappa=0.1$, and $F_z/(3\rho\vaz^3)=$
(a) 30, (b) 200, and (c) 1000.  Gas pressure has been neglected
and the vertical entropy gradient has been set to zero (i.e. $N_{\rm r}=0$).
}
\end{figure} 

\vfill\eject

\begin{figure}
\figurenum{5}
\epsscale{0.7}
\plotone{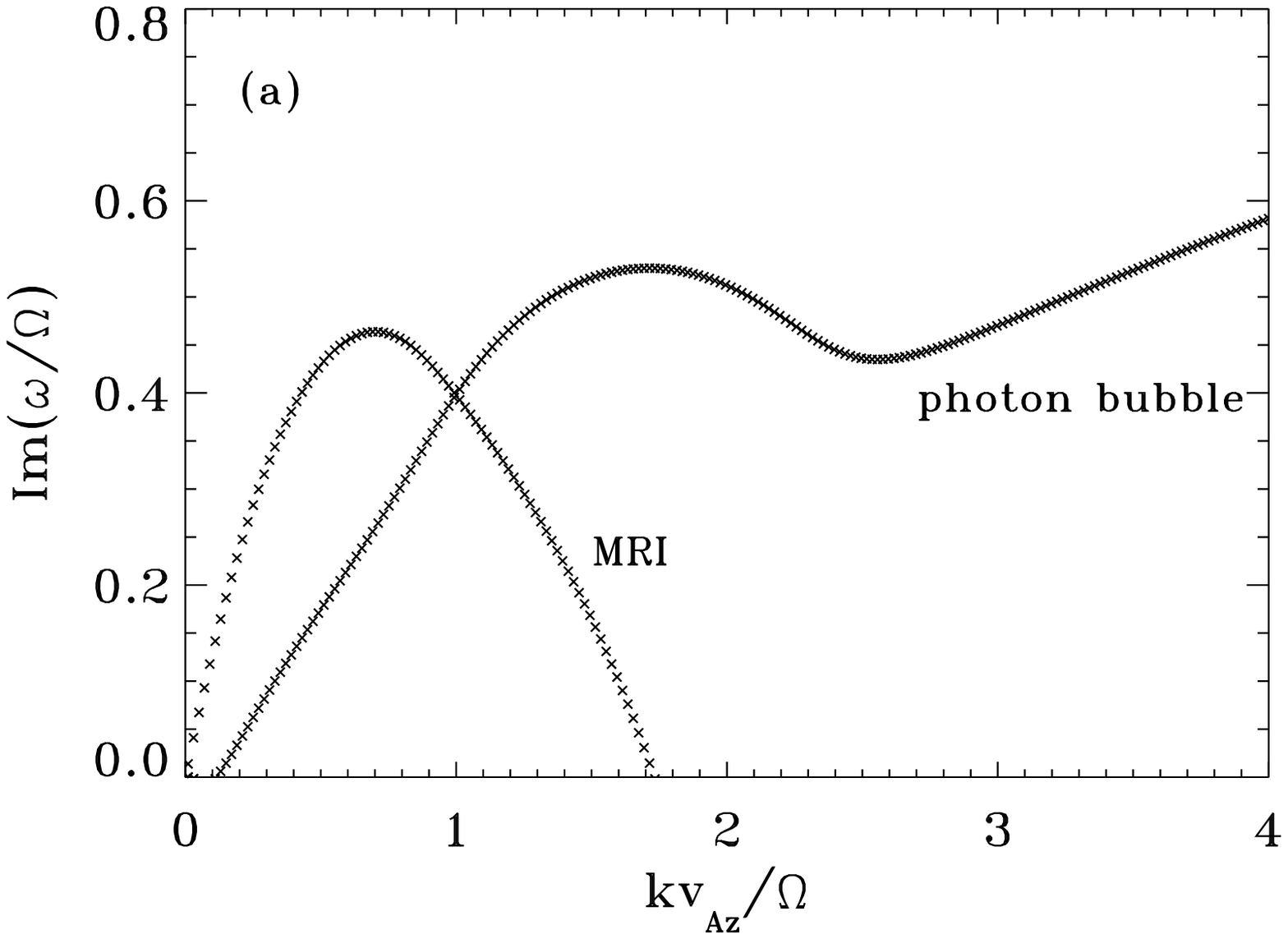}
\end{figure}

\begin{figure}   
\figurenum{5}
\epsscale{0.7}
\plotone{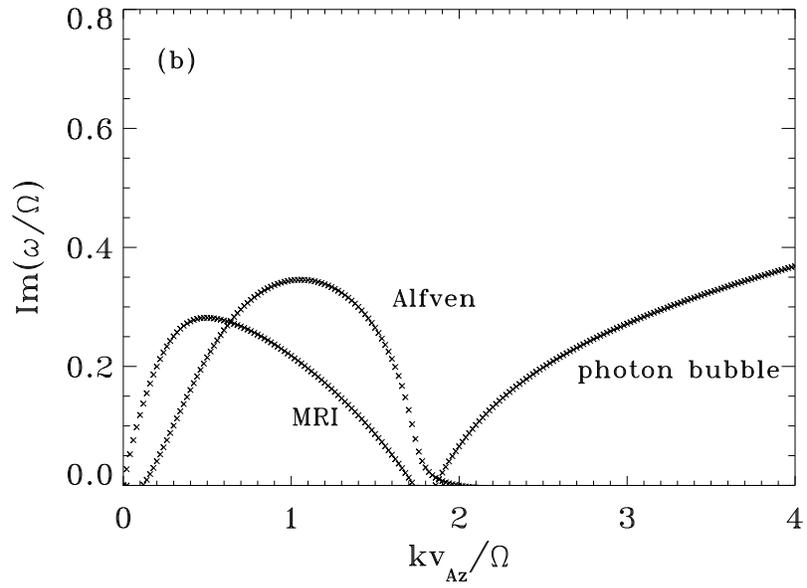}
\vskip0.1truein
\caption{Same as figure 4(b) only with a nonvertical magnetic field:
(a)$B_\phi=B_z$ and (b)$B_\phi=2B_z$.}
\end{figure} 

\end{document}